\documentstyle[12pt,epsfig]{article}
\begin{document}


\vspace{0.5cm}

\begin{center}
{\large \bf Limits on 
$\nu_{\mu}(\nu_{e}) \rightarrow \nu_{\tau}$  oscillations and  
tau neutrino magnetic moment from  neutrino-electron scattering} 
\end{center}
\vspace{0.5cm}
\begin{center}
{\large  
S.N.Gninenko\footnote{E-mail address: Sergei.Gninenko\char 64 cern.ch}}
\end{center}

\begin{center}
{ Institute for Nuclear Research, Russian Academy of Sciences,\\ 
Moscow 117 312, Russia}  
\end{center}

\begin{abstract}
The combined effect of neutrino flavor oscillations and neutrino magnetic 
moment on neutrino-electron scattering is discussed.\ It is shown, that
if the tau neutrino has a (large) magnetic moment and can oscillate into   
a neutrino of another flavor it can be used for a sensitive 
search for 
$\nu_{\mu}(\nu_{e}) \rightarrow \nu_{\tau}$ neutrino oscillations
and the tau neutrino magnetic moment in neutrino -electron scattering experiments.\ 
The combined limits on the mixing angles and the 
tau neutrino magnetic moment  
$sin^{2}2\theta_{\mu\tau}\times \mu_{\nu_{\tau}}^{2} < 1.1 \times 10^{-18}\mu_{B}^{2}$ and $sin^{2}2\theta_{e\tau}\times \mu_{\nu_{\tau}}^{2} < 2.3 \times 10^{-18}\mu_{B}^{2}$  for $\Delta m^{2} > 10~ eV^{2}$ are presented 
based on  
results of a study  of the neutrino-electron elastic scattering at LAMPF.\
For $\nu_{\mu}\rightarrow \nu_{\tau}$ oscillations, this would results in 
sensitive limit of $sin^{2}2\theta_{\mu\tau} < 4 \times 10^{-6}$, assuming the tau neutrino
magnetic moment being 
equal to the present experimental limit of $5.4 \times 10^{-7} \mu_{B}$.\
The tau neutrino magnetic moment would be  constrained to
 $\mu_{\nu_{\tau}} < 1.0 \times 10^{-8} \mu_{B}$, assuming the 
existence of $\nu_{\mu}\rightarrow \nu_{\tau}$ oscillations with the 
 mixing angle $sin^{2}2\theta_{\mu\tau}$ 
equal to the present experimental limit of 0.01 for $\Delta m^{2} > 10~ eV^{2}$.\
 Under 
similar assumptions the corresponding limits for $\nu_{e}\rightarrow \nu_{\tau}$ oscillations could be set to $sin^{2}2\theta_{e\tau} < 8 \times 10^{-6}$ and 
$\mu_{\nu_{\tau}} < 3.9 \times 10^{-9}\mu_{B}$.  
\end{abstract}

\vspace{1.0cm}

In the Standard Model(SM) the properties of neutrino are strictly  defined: it is a
massless particle with zero electric charge and magnetic moment.\ Beyond the SM
a nonzero mass of the neutrino is required for Grand Unified Theories
(for review see e.g. \cite{1}).\ In addition, massive tau neutrinos are natural candidates for the hot component of
the dark matter of the Universe.\ It is generaly assumed that 
 massive neutrinos could manifest themselves through the effect of 
neutrino oscillations first postulated by Pontecorvo \cite {2} a long time ago.\
Nonzero electromagnetic properties of the neutrino have  also been discussed in many
extensions of 
the SM \cite{1}.\  In its simplest extension, for example, 
neutrinos may have masses and acquire a magnetic moment through 
radiative corrections \cite{3,4}:
\begin{equation}
\mu_{\nu_{i}} = \frac{3eG_{F}m_{\nu_{i}}}{8\sqrt{2} \pi^{2}} = 3.2\times 10^{-19}( m_{\nu_{i}}/1~eV) \mu_{B} 
\end{equation}  
where $G_{F}$ is the Fermi constant, $\mu_{\nu_{i}} (i = 1, 2, 3 )$ is the neutrino magnetic moment,
 $m_{\nu_{i}}$ is the Dirac neutrino mass eigenstate and $\mu_{B} = e/2m_{e}$ is the Bohr magneton.\ This model, however predicts neutrino magnetic moments
 which are too small
to give any observable effects discussed in the present paper.\  
Some models predict for the neutrino 
magnetic moment values as large as $10^{-10}\mu_{B}$, see e.g. ref.\cite{5}, 
 which are large enough to be observed 
through neutrino  electromagnetic interactions.\

At present the $\nu_{\mu}\rightarrow \nu_{\tau}$ and $\nu_{e}\rightarrow \nu_{\tau}$ neutrino oscillations are a subject of a search at the  
CERN SPS wide-band neutrino beam by the CHORUS and NOMAD experiments \cite{6,7}.\ 
These searches are based on the appearance of tau neutrinos in a beam of 
muon neutrinos and requires detection of the $\nu_{\tau}$ charged current
 interaction

\begin{equation}
\nu_{\tau} N \rightarrow \tau^{-}X
\end{equation}  
The intrinsic fraction of $\nu_{\tau}$'s in the beam is expected to be 
negligibly small ($\sim 10^{-6}$).\
Two methods are used to detect  reaction (2).\ The CHORUS 
collaboration adopted the method  based on the direct observation the $\tau$ decay 
topology.\ In the NOMAD experiment a search of an apparent non-conservation 
of transverse momentum due to undetected neutrinos in  $\tau$ decays is used
\cite{8}.\

 In this note we show that combined existence  of  $\nu_{\mu}\rightarrow \nu_{\tau}$
and/or $\nu_{e}\rightarrow \nu_{\tau}$
 oscillations and  a (large) nonzero magnetic moment of the tau neutrino would 
increase 
the total rate of events in $\nu_{\mu}(\nu_{e})$ neutrino- electron scattering experiments.\
 This effect could be used for a sensitive  
search for these combined properties of the tau neutrino  in the reactions of neutrino - electron scattering.\


Assuming 
 that a muon neutrino beam has a component of tau neutrinos due to 
$\nu_{\mu}\rightarrow \nu_{\tau}$ oscillations.\ Then if magnetic moment of the
 $\nu_{\tau}$ exists,
it will contribute to a non-coherent part
of the $\nu_{\tau}e^{-}$ scattering cross section 
via the reaction that change the helicity of the tau neutrino (hence 
 right-handed neutrino states should exist) \cite{9}.\ 
 This 
might result in observable deviations from purely  $\nu_{\mu}e^{-}$ 
electro-weak reaction which is well predicted by the SM.\
  Indeed, since the 
electromagnetic cross section is orders of 
magnitude larger than the weak cross section, even a small fraction of 
tau neutrinos with nonzero magnetic moment in the muon neutrino beam could lead
 to an observable effects in $\nu_{\mu}e^{-}$  scattering,
 while the magnetic moment 
of muon neutrino could be small enough to contribute effectively to 
$\nu_{\mu}e^{-}$ scattering.\

The neutrino-electron scattering process via magnetic moment has a cross 
section \cite {9}:

\begin{equation}
\frac{d \sigma_{\mu}}{dy} = \frac{\pi \alpha^{2}}{m_{e}^{2}} 
\frac{\mu_{\nu}^{2}}{\mu_{B}^{2}} \frac{1-y}{y} 
\end{equation}

where $y = E_{e}/E_{\nu}$,  and 
$\mu_{\nu}$ is the neutrino magnetic moment.\ 
 The production rate of isolated electrons via 
$\nu_{\tau} e^{-}$  scattering in 
the detector depends also on the probability to find a $\nu_{\tau}$ neutrino in the 
neutrino beam.\ This probability can be calculated from the neutrino 
survival and transition probabilities.

 Consider first the case of a mixture of three massive neutrinos
\begin{equation}
\nu_{l}=\sum_{i=1}^{3}U_{li}\nu_{i},
\end{equation}
Here $\nu_{l},\nu_{i}$ are the fields of the neutrino with flavor $l$ and 
mass $m_{i}$, respectively  and $U$ is a 
unitary mixing  matrix.\  Ignoring possible $CP$ violation, for the probability of the transition  $\nu_{l} \rightarrow \nu_{l'}$ we have\cite {10}
\begin{equation}
P(\nu_{l} \rightarrow \nu_{l'})=\Bigl|
\delta_{ll'} - 4\sum_{i>j}U_{l'i}U_{l'j}U_{li}U_{lj}\times
 sin^{2}\frac{ \Delta m^{2}_{ij} L}{4E_{\nu}}\Bigr|
\end{equation}
where  $l,l' = e, \mu, \tau$, $L$ is the distance between the detector 
and the neutrino source and $\Delta m^{2}_{ij} = \bigl|m^{2}_{i} - m^{2}_{j}
\bigr|$.
 
In case of two-neutrino mixing, e.g. $\nu_{\mu} \rightarrow \nu_{\tau}$,
 neutrino states  evolve with a time $t$ as  

\begin{equation}
|\nu> (t) = a(t)|\nu_{\mu}>  + b(t)|\nu_{\tau}>
\end{equation}    

where $|\nu_{\mu}>$ and $|\nu_{\tau}>$ denote  weak eigenstates 
of $\nu_{\mu}$ and $\nu_{\tau}$ neutrinos, and $a^{2}(t),~b^{2}(t)$ are the 
probabilities to find $\nu_{\mu}$ or $\nu_{\tau}$ in the beam at a given moment
 $t$, respectively.\ It is assumed that $a^{2}(0) =1$ at $t=0$.\  
The probability $b^{2}(t)$ 
depends on the parameters of $\nu_{\mu} - \nu_{\tau}$ oscillations as :   

\begin{equation}
b^{2}(t\simeq \frac{L}{c}) = P(\nu_{\mu}\rightarrow\nu_{\tau}) = 
sin^{2}2\theta_{\mu\tau} sin^{2}\frac{\Delta m^{2} L}{4E} 
\end{equation}
or
\begin{equation}
P(\nu_{\mu}\rightarrow \nu_{\tau}) \approx sin^{2}2\theta_{\mu\tau} sin^{2}\frac{1.27 \Delta m^{2}(eV^{2}) L(km)}{E(GeV)}
\end{equation}

where $sin^{2}2\theta_{\mu\tau}$ is the
 mixing angle, and 
$\Delta m^{2} = \Delta m^{2}_{32}$.\

Now,
if an experiment sets a limit on a magnetic moment of the
neutrino $\mu_{\nu} < \mu_{0}$,  we may write in the general form
the following equation, taking into account  neutrino oscillations
between different neutrino flavors and nonzero magnetic moment
of the neutrinos:

\begin{equation}
 \sum_{l,l'=e,\mu,\tau}f_{\nu_{l}}\times [1 - P(\nu_{l} \rightarrow \nu_{l'})]\times
\mu_{\nu_{l}}^{2}  +  
f_{\nu_{l}}\times P(\nu_{l} \rightarrow \nu_{l'}) \times \mu_{\nu_{l'}}^{2} \leq 
\mu_{0}^{2}
\end{equation}

where $f_{\nu_{l}}$ is a factor corresponding to the flux of the 
$\nu_{l}$-neutrino at the source, and the fluxes are normalized to 1 = $\sum f_{\nu_{l}}$.\ Here we assume that   
detection efficiencies for  electrons produced in $\nu_{\l}e^{-}$ 
 scattering are the same for neutrino of all three flavors.\ Note, that limit (9)
is set based on an agreement between observed cross section 
in $\nu e^{-}$ scattering experiment and SM expectations.\ The 
$\nu_{\mu} \rightarrow \nu_{e}$ oscillations could also affect the observed 
$\nu e$ cross section because of the difference in cross sections for 
$\nu_{e} e^{-}$ and $\nu_{\mu}(\nu_{\tau})e^{-}$ scattering.\ However, 
 according to the present data on
$\nu_{\mu} \rightarrow \nu_{e}$ oscillation from the LSND experiment,  
 $P(\nu_{\mu} \rightarrow \nu_{e}) \approx 3 \times 10^{-3}$ for (large)  
 $\Delta m^{2} > 10~ eV^{2}$ \cite{11}. This could shift the   
$\nu_{\mu}e^{-}$ cross section by at most 1$\%$, which is small in 
comparison with the statistical and systematic errors of the experiments discussed below.\ More  detailed discussions of the effect of neutrino flavor oscillations
on $\nu e^{-}$ scattering  can be found in ref. \cite {12}.\ We will ignore 
this effect to avoid complication of our formula (9).

 The best limit on neutrino magnetic moment  
 was obtained from the experiment on study of the $\nu_{e}e^{-}$ elastic 
scattering at LAMPF, ref. \cite{13,14}, and was confirmed by the CHARM II experiment at CERN
at much higher energy, ref.\cite{15}.\ The average $L/E$ 
ratio was $\approx 0.3$ in ref.\cite{13,14}, and $\approx 0.05$ in ref.\cite{15}.\

 The upper limits on $\mu_{\nu_{\mu}}$ and $\mu_{\nu_{e}}$ from 
the LAMPF experiment was obtained using the following equation 
( ref.\cite{14}, Eq.(23))
\begin{equation}
\mu_{\nu_{e}}^{2} + 2.1\times \mu_{\nu_{\mu}}^{2}  \leq \mu_{0}^{2}=1.16 \times
10^{-18}\mu_{B}^{2}
\end{equation}
and therefore 
\begin{equation}
\mu_{\nu_{\mu}} < 7.4 \times 10^{-10} \mu_{B}~~~~~~(\mu_{\nu_{e}} = 0)
\end{equation}
\begin{equation}
\mu_{\nu_{e}} < 10.8 \times 10^{-10} \mu_{B}~~~~~~(\mu_{\nu_{\mu}} = 0)
\end{equation}
 
Taking into account $\nu_{\mu}(\nu_{e}) \rightarrow \nu_{\tau}$ 
oscillations  Eq.(10) can be rewritten in the form similar to Eq.(9)

\begin{eqnarray}
[1 - P(\nu_{e} \rightarrow \nu_{\tau})]
\times \mu_{\nu_{e}}^{2} + P(\nu_{e} \rightarrow \nu_{\tau}) \times 
\mu_{\nu_{\tau}}^{2}\nonumber~~~~~~~~~~~~\\
 +~2.1\times [1 - P(\nu_{\mu} \rightarrow \nu_{\tau})] \times \mu_{\nu_{\mu}}^{2} +  
2.1\times P(\nu_{\mu} \rightarrow \nu_{\tau}) \times \mu_{\nu_{\tau}}^{2} \leq 
\mu_{0}^{2}
\end{eqnarray}
which can be used to obtain combined limits on the probability of
 $\nu_{\mu}(\nu_{e}) \rightarrow \nu_{\tau}$ oscillations versus  
tau neutrino magnetic moment:
  
\begin{equation}
P(\nu_{e}\rightarrow \nu_{\tau})\times \mu_{\nu_{\tau}}^{2} + 2.1 \times 
P(\nu_{\mu}\rightarrow \nu_{\tau})\times \mu_{\nu_{\tau}}^{2} \leq 1.16 \times 
10^{-18}\mu_{B}^{2}~~~~(\mu_{\nu_{e}}=\mu_{\nu_{\mu}}=0)
\end{equation}
\begin{equation}
P(\nu_{e}\rightarrow \nu_{\tau})\times \mu_{\nu_{\tau}}^{2}  \leq 1.16 \times 
10^{-18}\mu_{B}^{2}~~~~~~~~(P(\nu_{\mu}\rightarrow \nu_{\tau})=0,~\mu_{\nu_{e}}=\mu_{\nu_{\mu}}=0)
\end{equation}
\begin{equation}
P(\nu_{\mu}\rightarrow \nu_{\tau})\times \mu_{\nu_{\tau}}^{2} \leq 5.5 \times 
10^{-19}\mu_{B}^{2}~~~~~~~~( P(\nu_{e}\rightarrow \nu_{\tau})=0,~\mu_{\nu_{e}}=\mu_{\nu_{\mu}}=0)
\end{equation}

From Eqs.(15,16) combined limits on the mixing angles and tau neutrino 
magnetic 
moment  for  $\Delta m^{2} > 10~ eV^{2}$, which corresponds to the large
$\Delta m^{2}$ region in the LAMPF experiment can be derived

\begin{equation}
sin^{2}2\theta_{e\tau}\times \mu_{\nu_{\tau}}^{2} \leq 2.3 \times 10^{-18}\mu_{B}^{2} 
\end{equation}
\begin{equation}
sin^{2}2\theta_{\mu\tau}\times \mu_{\nu_{\tau}}^{2} \leq 1.1 \times 10^{-18}\mu_{B}^{2} 
\end{equation}

Assuming the tau-neutrino magnetic moment to be equal to the 
present experimental limit of $5.4 \times 10^{-7} \mu_{B}$ from the BEBC experiment at CERN,
ref.\cite{16}, the following 
limits on the
oscillation probabilities $P(\nu_{e} \rightarrow \nu_{\tau})$, $P(\nu_{\mu} \rightarrow \nu_{\tau})$ 
and mixing angles $sin^{2}2\theta_{e\tau}$, $sin^{2}2\theta_{\mu\tau}$   can be obtained from Eqs.(15,16,17,18), respectively
\begin{equation}
P(\nu_{e} \rightarrow \nu_{\tau}) < 4 \times 10^{-6}
\end{equation}
\begin{equation}
sin^{2}2\theta_{e\tau} < 8 \times 10^{-6}
\end{equation}
\begin{equation}
P(\nu_{\mu} \rightarrow \nu_{\tau}) < 2 \times 10^{-6}
\end{equation}
\begin{equation}
sin^{2}2\theta_{\mu\tau} < 4 \times 10^{-6}
\end{equation}

Under assumption of the existence of neutrino
oscillations with  mixing angles  equal
 to the present  experimental limits of 
$sin^{2}2\theta_{e\tau} = 0.15$ obtained from   
$\overline{\nu_{e}}\rightarrow \nu_{x}$ disappearance reactor experiments 
in  Bugey \cite{17} and in Krasnoyarsk \cite{18},
 or $sin^{2}2\theta_{\mu\tau} = 0.01$
for $\nu_{\mu}\rightarrow \nu_{\tau}$, 
obtained from the exclusion plot of ref.\cite{19} for $\Delta m^{2} > 10~ eV^{2}$,  the following upper 
bounds on the magnetic moment of the tau neutrino can be derived from
Eqs.(17,18):
\begin{equation}
\mu_{\nu_{\tau}} < 3.9 \times 10^{-9}\mu_{B}~~~~~(\nu_{e}\rightarrow \nu_{\tau})
\end{equation}
\begin{equation}
\mu_{\nu_{\tau}} < 1.0 \times 10^{-8}\mu_{B}~~~~~(\nu_{\mu}\rightarrow \nu_{\tau}) 
\end{equation}

In (23), no distinction has been made between $\nu_{e}$ and 
$\overline{\nu_{e}}$.\
 The exclusion regions in the    
($sin^{2}2\theta_{e\tau},~ \mu_{\nu_{\tau}}/\mu_{B}$) and 
($sin^{2}2\theta_{\mu\tau},~ \mu_{\nu_{\tau}}/\mu_{B}$) planes 
obtained from Eqs.(17,18) are 
shown in Figure 1,2 , respectively.\ Experimentally excluded regions for
$\mu_{\nu_{\tau}}$, 
$sin^{2}2\theta_{e\tau}$ and
$sin^{2}2\theta_{\mu\tau}$ are also indicated.\\

Assuming the existence of $\nu_{\mu}\rightarrow \nu_{\tau}$ neutrino
oscillations with  mixing angles $sin^{2}2\theta_{\mu\tau}$ equal
 to the present  experimental limits of 
$3.5 \times 10^{-3}$ from the CHORUS \cite{20}, or
 $3.7 \times 10^{-3}$ from the NOMAD \cite{21}
experiments, the limit of
\begin{equation}
\mu_{\nu_{\tau}} < 1.8 \times 10^{-8}\mu_{B}~~~~~(\nu_{\mu}\rightarrow \nu_{\tau}) 
\end{equation}
can also be obtained, however for the mass region $\Delta m^{2} > 1000~eV^{2}$.

Using the upper limit on muon neutrino magnetic moment 
from the CHARM II experiment \cite{15} one can obtain limits on 
$P(\nu_{\mu} \rightarrow \nu_{\tau})$, $sin^{2}2\theta_{\mu\tau}$ and 
$\mu_{\nu_{\tau}}$ of the same
order as the limits (21),(22), and (24), respectively.\
Note that 
 to separate an oscillation signal from magnetic moment scattering 
in $\nu e^{-}$ reaction one can 
look  at its energy or distance dependence.\ For example, the probability of oscillations  
changes as a $E_{\nu}^{-2}$ for small $\Delta m^{2}$, while the integral cross 
section for the magnetic moment scattering  varies only logarithmically 
with $E_{\nu}$.\ 

\vspace{0.5cm}

{\large \bf Acknowledgements}\\

The author wishes to thank L.~Camilleri, L.~DiLella, E.~do~Couto e Silva, 
V.A.~Kuzmin and A.~Rubbia 
for useful discussions and comments. Communications with K.~Winter 
on latest CHORUS results are gratefully acknowledged.\

\newpage
 \begin{figure}[hbt]
   \mbox{\epsfig{file=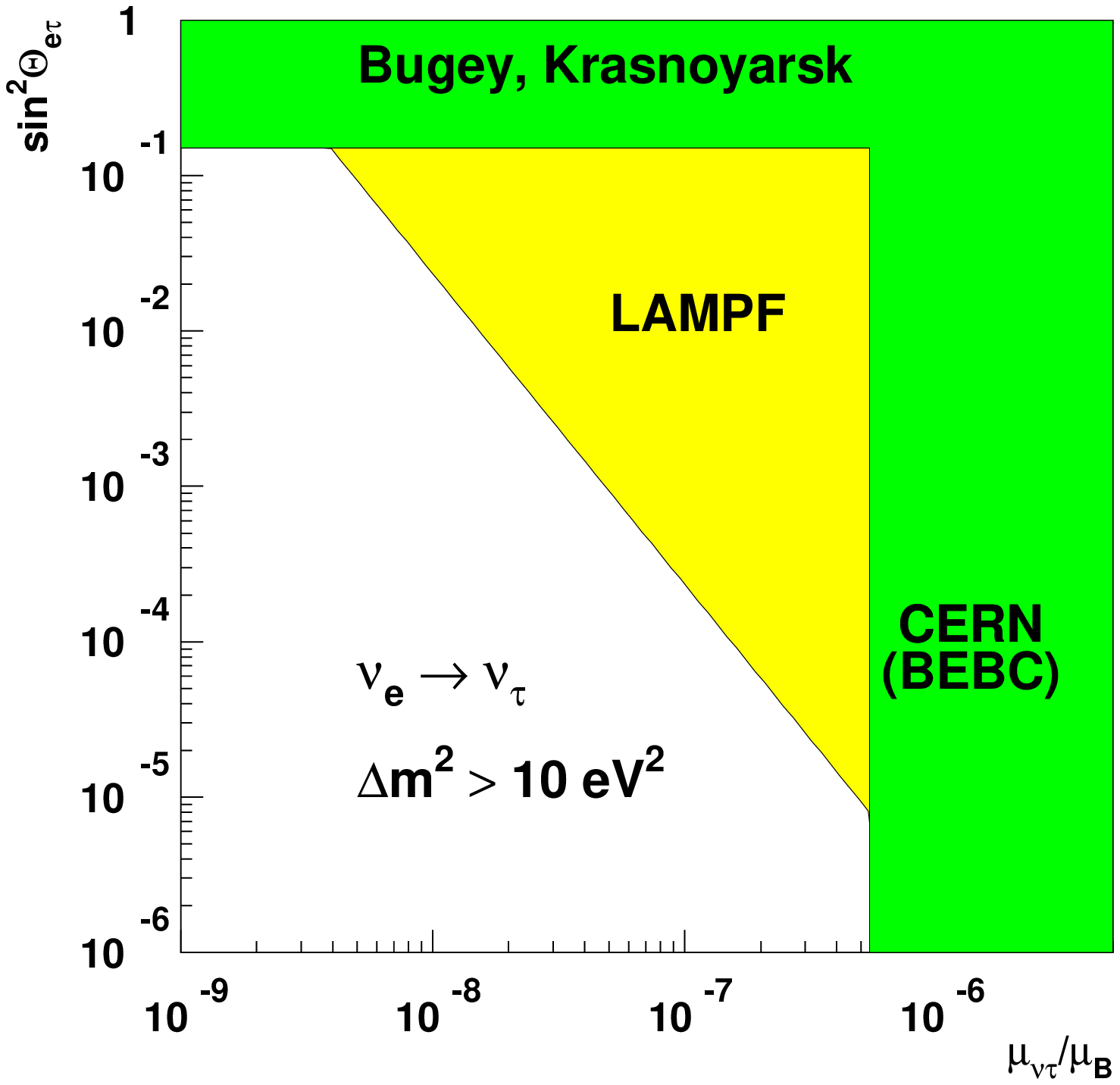,height=140mm}}
    \centering
  \caption{\em The exclusion regions for combination of   
 $\nu_{e} \rightarrow \nu_{\tau}$ oscillations and 
tau neutrino magnetic moment $\mu_{\nu_{\tau}}$  in the
($sin^{2}2\theta_{e\tau},~ \mu_{\nu_{\tau}}/\mu_{B}$) plane.\
Light shaded area is excluded by this analysis based on the LAMPF results  
on $\nu e^{-}$ scattering  \cite {13,14}.\ Dark shaded area 
corresponds to the experimentally excluded regions 
$\mu_{\nu_{\tau}}> 5.4 \times 10^{-7} \mu_{B}$ \cite{16} and 
$sin^{2}2\theta_{e\tau} > 0.15$  \cite{17,18}.
 The limits are valid for
$\Delta m^{2} > 10~ eV^{2}$. 
} 
  \label{figure 2:}
\end{figure}

\newpage
 \begin{figure}[hbt]
   \mbox{\epsfig{file=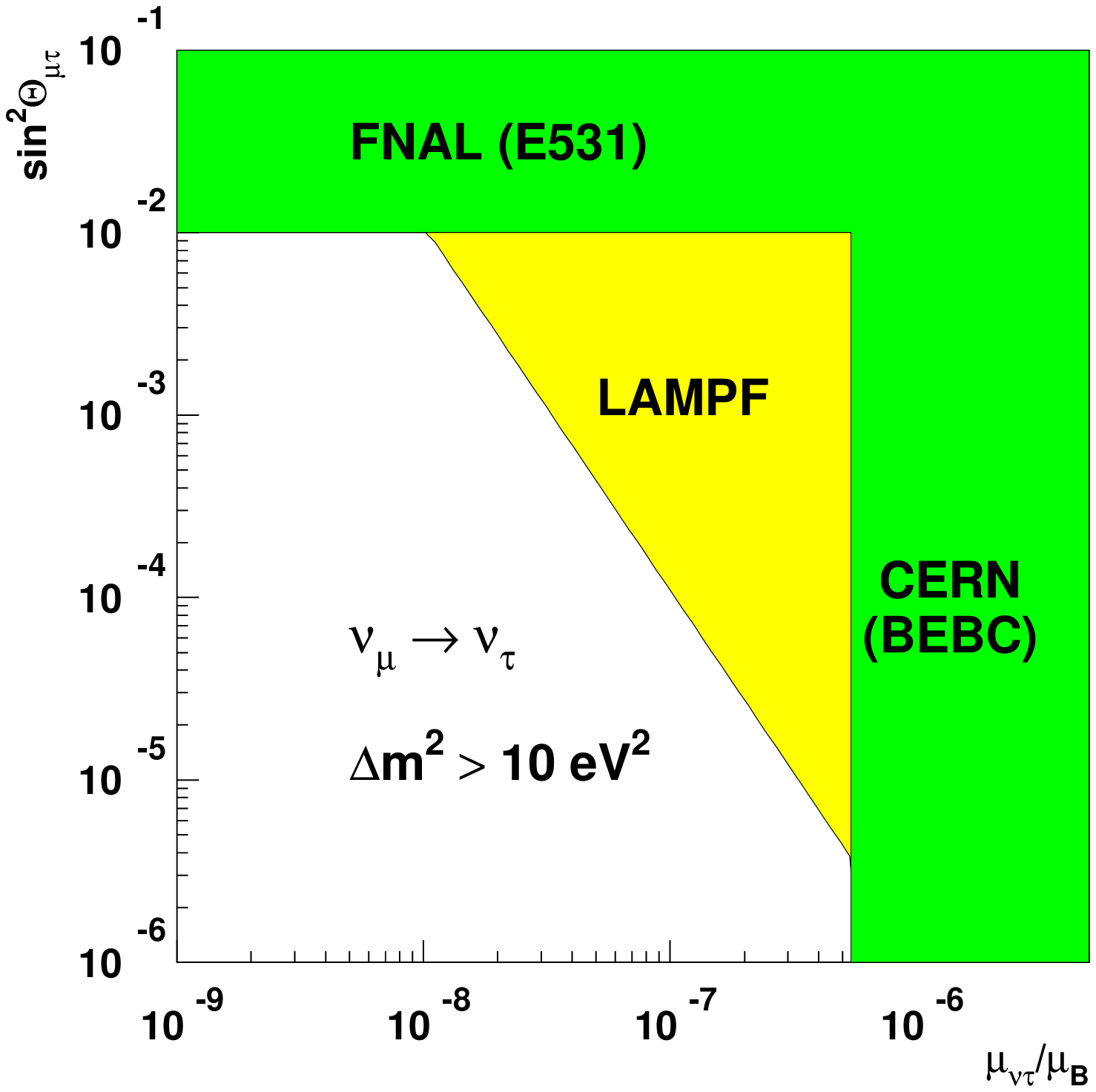,height=180mm}}
    \centering
  \caption{\em The exclusion regions  for combination of
 $\nu_{\mu} \rightarrow \nu_{\tau}$ oscillations and 
tau neutrino magnetic moment $\mu_{\nu_{\tau}}$   in the
($sin^{2}2\theta_{\mu\tau},~ \mu_{\nu_{\tau}}/\mu_{B}$) plane.\
Light shaded area is excluded by this analysis based on the LAMPF results 
on $\nu e^{-}$ scattering \cite {13,14}.\ Dark shaded area 
corresponds to the experimentally excluded regions 
$\mu_{\nu_{\tau}}> 5.4 \times 10^{-7} \mu_{B}$ \cite{16} and 
$sin^{2}2\theta_{\mu\tau} > 0.01$  \cite{19}.
 The limits are valid for
$\Delta m^{2} > 10~ eV^{2}$.
} 
  \label{figure 1:}
\end{figure}

\end{document}